\pdfoutput=1
\documentclass[lettersize,journal]{IEEEtran}
\usepackage{lipsum}
\usepackage{hyperref}
\hypersetup{
    colorlinks=true,
    linkcolor={red!50!black},
    citecolor={blue!65!black},
    urlcolor={blue!80!black}
}
\usepackage{amsmath}
\usepackage{subfig}
\usepackage{wrapfig}
\usepackage{xcolor}
\usepackage{algorithm,algpseudocode}

\usepackage{multirow}
\usepackage{makecell}
\usepackage{soul}
\usepackage{amssymb}
\usepackage{placeins}
\usepackage[normalem]{ulem}
\usepackage{amsthm}
\newtheorem{proposition}{Proposition}
\usepackage{graphicx}
\newcounter{panel}

\usepackage{caption}
\usepackage{subcaption}

\begin{document}
\title{Quantifying Membership Inference Risks in Genomic Data Sharing: The Role of Attacker's Auxiliary Information}
\author{Maryam Ghasemian, Lynette Hammond Gerido, Erman Ayday$^{\ast}$%
\thanks{Maryam Ghasemian and Erman Ayday are with the Department of Computer and Data Sciences , Case School of Engineering, Case Western Reserve University, Cleveland, OH, USA. Email: \{maryam.ghasemian, erman.ayday\}@case.edu
}%
\thanks{Lynette Hammond Gerido is with the Department of Bioethics, School of Medicine, Case Western Reserve University, Cleveland, OH, USA. Email: LHGerido@case.edu}%
\thanks{$^{\ast}$Corresponding author: Erman Ayday (e-mail: erman.ayday@case.edu).}}
\maketitle
\begin{abstract}
Genomic data sharing is essential for biomedical research but introduces significant privacy risks, including membership inference attacks, in which an adversary determines whether a specific individual's data was used in a study. Existing privacy protections are typically applied after a release format has already been chosen, leaving researchers without a principled method to assess dataset-specific risk beforehand. We propose a server-assisted, principal component analysis-based pre-release auditing framework that maps genomic data into a shared low-dimensional space, applies epsilon-local differential privacy via the Laplace mechanism with a formally derived sensitivity for principal component analysis scores, and estimates membership inference attack power at a controlled 5\% false-positive rate using a distance-threshold attack. A membership-neutrality diagnostic based on Unbiased Reconstruction Robustness is reported alongside attack power, precision, and recall. Experiments using the openSNP and 1000 Genomes Project datasets across multiple dimensionalities and privacy budgets show that separation between the researcher's dataset and public auxiliary data in principal component analysis space is a dominant determinant of attack success, exceeding the effect of perturbation level alone. Comparison against a genome-wide association study-style likelihood ratio test baseline further contextualizes the measured risk. This framework enables researchers, data repositories, and institutional review bodies to quantify dataset-specific membership inference risk before genomic data publication, supporting more informed and privacy-aware biomedical data-sharing decisions.
\end{abstract}


\vspace{-10pt}
\section{Introduction}\label{sec:introduction}
The increasing availability of large-scale genomic datasets has fueled transformative advances in personalized medicine, population genetics, and biomedical research. However, the widespread sharing of genomic data introduces significant privacy risks. Genomic information is inherently sensitive and uniquely identifiable, making even partial exposure potentially irreversible. Among the most critical threats are membership inference attacks (MIAs), where adversaries attempt to determine whether a specific individual’s data is part of a shared dataset. Such privacy breaches pose a significant threat to individual confidentiality and may hinder the willingness of individuals to contribute to genomic research initiatives and undermine public trust.

Membership inference risks in genomics are well established. Homer et al. \cite{Homer2008Resolving} demonstrated that individuals could be detected in aggregate GWAS allele-frequency data, prompting major repositories to restrict access to aggregate genomic data~\cite{Zerhouni2008Protecting, Couzin2008Whole-Genome}. Subsequent work has shown that leakage can occur across multiple release formats, including linkage disequilibrium statistics~\cite{Wang2009Learning}, haplotype summaries~\cite{Bu2021Haplotype}, and familial SNP data~\cite{humbert2017Quantifying}. These findings motivate treating privacy risk as a pre-release data-sharing concern rather than a post-release mitigation problem.

This issue has become more pressing as large public and controlled-access genomic repositories continue to grow. Resources such as dbGaP~\cite{dbGap} and the UK Biobank~\cite{UK-BioBank} provide increasingly rich auxiliary information that can strengthen membership inference. Existing privacy-preserving methods--including differential privacy mechanisms for GWAS statistics~\cite{chen2021Differential,Halimi2022Privacy}, dependent local DP methods for genomic data~\cite{Yilmaz2022Genomic}, and game-theoretic noise calibration strategies~\cite{zhang2024bayesnash}--focus primarily on how to protect a release once a release mechanism has been selected. Much less attention has been given to the earlier question that data stewards, repositories, and investigators often face in practice: \textit{before a release is made, how much membership risk is likely under realistic auxiliary-information conditions?}

This work addresses that question through a pre-release privacy auditing framework for genomic data sharing. The key idea is to assess membership risk in a shared low-dimensional representation space before publication. Principal Component Analysis (PCA) is used not only to reduce dimensionality, but also to map high-dimensional genomic profiles into a lower-dimensional geometry in which Euclidean distances between case, control, and target groups are more interpretable than distances in the original SNP space. Within that shared PCA space, the spatial relationship between the researcher's dataset and public auxiliary datasets can be used to quantify the success of a distance-threshold-based MIA at a controlled false-positive rate. Under this attack model, greater geometric separation makes membership inference easier, whereas greater overlap reduces separability between members and non-members.

To instantiate this workflow, the researcher computes PCA locally and perturbs the resulting score vectors with Laplace noise before sharing them for risk assessment. In our implementation, the perturbation scale is set to $\lambda = 2\sqrt{d}/\epsilon$ for PCA dimensionality $d$ and perturbation level $\epsilon$. The server then projects public reference datasets into the same PCA coordinate system and estimates MIA risk by comparing geometric proximity patterns in that space. This use of perturbation should be interpreted
carefully. Because the PCA model itself is shared and is not privately learned in the current design, the contribution of this work is not a claim of end-to-end local differential privacy for the full release pipeline. Rather, we position the method as a pre-release privacy risk auditing workflow that helps researchers and governance bodies evaluate genomic sharing decisions under
realistic auxiliary-information scenarios.

Risk estimation is performed by a semi-trusted server rather than by the researcher alone. This design is motivated by practical repository and governance considerations. First, the server can maintain up-to-date snapshots of public genomic resources that individual research groups may not have the capacity or permissions to curate locally. Second, an independent service can provide a reproducible and auditable risk assessment that may be referenced in
institutional review, repository governance, or data-access decision-making. Third, a shared service can support more standardized evaluation across studies that are released into overlapping public reference ecosystems. In this sense, the proposed workflow is intended not only as a technical method, but also as a decision-support component for more trustworthy genomic data sharing.

The main contributions of this paper are as follows:

\begin{enumerate}

  \item \textbf{A Pre-Release Privacy Auditing Framework:} We introduce a server-assisted, PCA-based workflow for estimating membership inference risk before genomic research outcomes are published. The framework is designed to support release decisions in settings where auxiliary genomic data may be available to adversaries.


  \item \textbf{Geometric Risk Characterization in PCA Space:} Using openSNP and the 1000~Genomes Project, we show that geometric separation between the researcher's dataset and public reference datasets is a dominant determinant of MIA success under the studied distance-threshold attack. Greater
  separation increases measured attack power, whereas greater overlap reduces separability.

  \item \textbf{Membership-Neutrality Diagnostic:} We report U-ReRo as a secondary diagnostic to assess whether the applied perturbation affects members and non-members asymmetrically, complementing standard MIA power, precision, and recall metrics.

  \item \textbf{Governance-Oriented Release Support:} We compare the proposed auditing workflow against a GWAS-style LRT baseline and present a decision process for interpreting quantified MIA risk in pre-release genomic data-sharing settings.

\end{enumerate}

The remainder of this paper is organized as follows.
Section~\ref{sec:background} provides background on PCA, perturbation, and evaluation metrics.
Section~\ref{sec:system} describes the system and threat model.
Section~\ref{sec:framework} details the proposed framework.
Section~\ref{sec:evaluation} reports the empirical evaluation.
Section~\ref{sec:discuss} discusses findings and implications.
Section~\ref{sec:conclusion} concludes.

\begin{table*}[!t]
\centering
\caption{Table of symbols and notations. }
\label{tab:symbols_notations}
\begin{tabular}{|l|l|}
\hline
\textbf{Symbol} & \textbf{Description}  \\
\hline
$D_R$ & Researcher's Dataset \\
$ND_R$ & Researcher's noisy data produced as a result of Laplace noise\\
$D_p$ & Publicly available dataset\\
MIA & Membership Inference Attack\\
$\epsilon$ , eps & Privacy Parameter \\
k & Number of clusters \\
$d$ & Number of Principal Components\\
U-ReRo & Unbiased Reconstruction Robustness\\
$\eta$ & Error threshold in U-ReRo \\
$\gamma$ & Reconstruction probability difference threshold in U-ReRo\\
\hline
\end{tabular} 
\end{table*}
\section{Background}\label{sec:background}
In this section we review some background and definitions of the local differential privacy. Extended background and related work are provided in Appendix~B.
\subsection{Principal Component Analysis (PCA)}
PCA is the standard dimensionality reduction tool in genomic analysis. Given a high-dimensional dataset of SNP profiles, PCA computes the eigenvectors of the sample covariance matrix and projects each individual's genotype vector onto the top-$d$ eigenvectors, yielding a compact set of uncorrelated scores called principal components. The resulting components are ordered by the amount of variance they explain, so that the first few dimensions capture the dominant axes of variation in the data. In genomics, these axes typically reflect population structure: the top components separate individuals by ancestry and geographic origin, making PCA a natural space in which to assess geometric relationships between datasets from different populations (\cite{Abdi2010Principal, Pedregosa2011Scikit}).

\subsection{Local Differential Privacy and Sensitivity}
Local Differential Privacy (LDP) ~\cite{costello2015geppetto,davidson2011on, Duchi2013Local, Kairouz2014Extremal} is a more restricted form of traditional differential privacy ~\cite{Dwork2008Differential, davidson2011provenance}. Unlike traditional differential privacy, LDP does not rely on a trusted third party and provides a higher level of data protection for users. In LDP, each user modifies their own data before sharing them with a data aggregator. The aggregator only sees the perturbed data, ensuring privacy. 
An algorithm \(A\) satisfies $\epsilon$-local differential privacy ($\epsilon$-LDP) if, for any input values $v_1$ and $v_2$ and any output $y$: 
\begin{math}
  Pr[A(v_1)=y]\leq e^\epsilon Pr[A(v_2)=y]  
\end{math},
The Laplace mechanism~\cite{Dwork2006Calibrating} is used to instantiate LDP.
For a numerical function $f$ with $\ell_1$ sensitivity $s$, the mechanism
$F(\mathbf{x}) = f(\mathbf{x}) + \text{Lap}(s/\epsilon)$ satisfies
$\epsilon$-LDP.

A key technical requirement is the derivation of the $\ell_1$ sensitivity of
the PCA score function, which determines the noise scale $\lambda = s/\epsilon$.
In prior work~\cite{Yilmaz2022Genomic, chen2021Differential}, sensitivity was either stated without derivation or derived only for simpler functions. The following proposition
provides the analytical derivation for the PCA score function used in this framework.

\begin{proposition}[\textbf{$\ell_1$ Sensitivity of PCA Scores}]
\label{prop:sensitivity}
For the fixed PCA score function $f(\mathbf{x}_i) = (\mathbf{x}_i - \boldsymbol{\mu})\mathbf{V}_{1:d}$, where normalized genotype vectors satisfy $\mathbf{x}_i \in [0,1]^p$, The $\ell_1$ sensitivity is $2\sqrt{d}$. Therefore, the Laplace scale used in this framework is $\lambda = 2\sqrt{d}/\epsilon$. This dependence on $\sqrt{d}$ motivates using a small number of principal components.
\end{proposition}
The full proof is provided in Appendix~A. 





\subsection{Unbiased Reconstruction Robustness (U-ReRo)}
\label{sec:urero}

Traditional metrics for evaluating MIA robustness in differential privacy often overlook biases in reconstruction accuracy for members versus non-members. U-ReRo~\cite{Guerra2024Analysis} is a membership-neutral metric that evaluates the impact of DP mechanisms equally for both groups. U-ReRo measures the difference in reconstruction probabilities under a defined error threshold $\eta$, ensuring this difference remains below a tolerance $\gamma$.

\textbf{Instantiation in the proposed framework.}
Reconstruction of individual $i$ is defined as successful if the $\ell_2$ distance between the reconstructed score $\hat{\mathbf{x}}_i = \tilde{\mathbf{y}}_i \cdot \mathbf{V}^{\dagger} + \boldsymbol{\mu}$ and the true genotype $\mathbf{x}_i$ is below threshold $\eta = 0.5$. The following quantities are computed:
\begin{align*}
P_{\text{member}}    &= \bigl|\bigl\{i \in D_R :
                       \|\hat{\mathbf{x}}_i - \mathbf{x}_i\|_2 \leq \eta\bigr\}\bigr|
                       \;/\; |D_R|, \\
P_{\text{nonmember}} &= \bigl|\bigl\{j \notin D_R :
                       \|\hat{\mathbf{x}}_j - \mathbf{x}_j\|_2 \leq \eta\bigr\}\bigr|
                       \;/\; |D_\text{ref}|, \\
\text{U-ReRo}        &= |P_{\text{member}} - P_{\text{nonmember}}|.
\end{align*}

When $\text{U-ReRo} \leq \gamma = 0.5$, the mechanism is considered
membership-neutral. In the evaluation reported in this work, U-ReRo
consistently remained $\ll 0.1$, substantially below this threshold.


\section{System and threat Model}\label{sec:system}
In this section, the system model and threat model are described. The approach considers practical scenarios involving real-world genomic datasets and research followers leveraging auxiliary information. Table~\ref{tab:symbols_notations} summarizes all symbols and notations used throughout the paper.
\subsection{System Model}
The system model consists of three key entities: the researcher, the server, and the research follower.

\textbf{\textit{Researcher}}:
The researcher collects a genomic dataset $D_R$, conducts research, and intends to publish research outcomes. To estimate the privacy risk associated with dissemination of research findings, the researcher shares metadata derived from $D_R$ with the server. To protect the privacy of research participants, the researcher (i) shares metadata rather than the raw research dataset, and (ii)
applies a differential privacy mechanism (Laplace noise) to the metadata before sharing with the server. Specifically, the researcher performs PCA locally to obtain score vectors $\mathbf{Y} = (\mathbf{X} - \boldsymbol{\mu})\mathbf{V}_{1:d}$, applies $\varepsilon$-LDP to produce $\tilde{\mathbf{Y}} = \mathbf{Y} +\text{Lap}(\lambda)$, and shares $\tilde{\mathbf{Y}}$ and the PCA model $(\mathbf{V}, \boldsymbol{\mu}, d)$ with the server. Raw genomic data is never transmitted.

\textbf{\textit{Server}}:  
The server receives the DP-noised scores $\tilde{\mathbf{Y}}$ and the PCA model from the researcher, and independently accesses relevant publicly available datasets $D_\text{ref}$. Using the shared PCA model, the server projects $D_\text{ref}$ into the same coordinate system as $\tilde{\mathbf{Y}}$ and computes distance-based metrics to estimate MIA risk. The server then reports this risk estimate to the researcher, enabling adjustment of privacy parameters prior to publication.

The server-assisted design supports up-to-date auxiliary-data curation, independent auditability, and standardized risk assessment across studies. It maintains continuously updated of multiple public genomic repositories, including the 1000~Genomes Project~\cite{1000genom}, dbGaP~\cite{dbGap}, and the UK Biobank~\cite{UK-BioBank}, that individual researchers typically do not maintain locally. Also, independent risk assessment provides institutional auditability. Additionally, the server can aggregate risk across studies, estimating how a new release interacts with prior releases using the same public datasets.

\textbf{\textit{Research Follower}}:  
The research follower is an entity that receives and uses published research outcomes. In adversarial scenarios, research follower may exploit publicly available genomic repositories--such as dbGaP~\cite{dbGap} or the UK Biobank~\cite{UK-BioBank}--to obtain auxiliary datasets and enhance membership inference capabilities. These repositories provide summary statistics, population-level data, or other reference information that can be incorporated into probabilistic or distance-based membership inference analyses.

\begin{figure*}[!t]%
   \centering
    \includegraphics[width=0.9\linewidth]{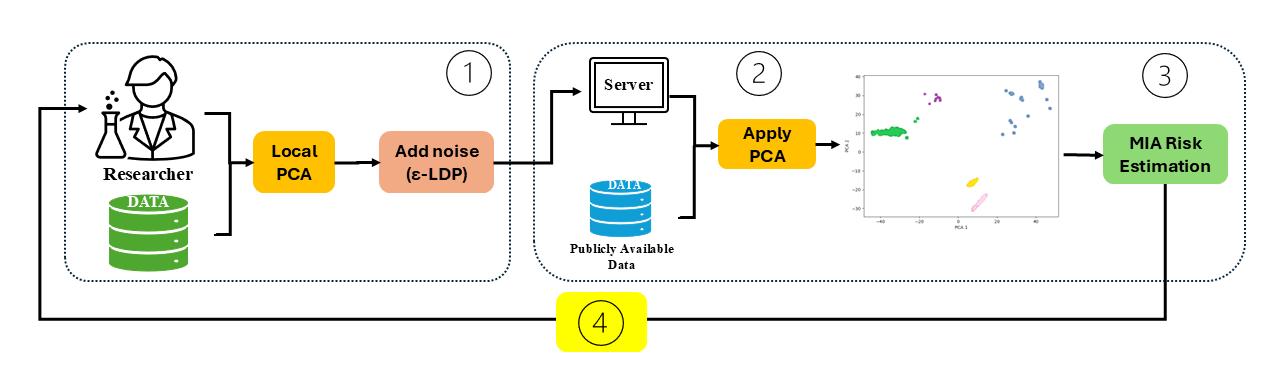}
    \caption{An overview of the proposed framework. (1) The researcher collects genomic data, performs local PCA, shares the PCA model and DP-noised scores with the server. (2) The server integrates public datasets and, using the shared PCA model, maps auxiliaries into the same coordinate system, and (3) The server estimates membership inference risk by analyzing the spatial relationships among the datasets. (4) The resulting risk estimation is reported back to the researcher, guiding them in fine-tuning privacy parameters prior to publishing research outcomes.
    }
    \label{fig:system_model}
\end{figure*}

\subsection{Threat Model}
In the adversarial setting, a research follower may leverage publicly available datasets, statistical summaries, or other auxiliary information to infer whether a specific individual is part of $D_R$. By proactively estimating MIA risks, the researcher and server aim to minimize the research follower's success probability before publication and support safer release decisions.

The primary threat in the proposed system arises from the semi-trusted but potentially compromised server. The server receives PCA-transformed representations derived from the researcher's DP-processed outputs and has access to the public reference datasets used for risk quantification. If the server deviates from its intended role, it could combine these sources to perform re-identification or membership-inference attacks. This risk is compounded by the fact that even low-dimensional projections can retain identifiable structure, particularly when cross-referenced with auxiliary information from public repositories.

Two mechanisms are used to address the server threat. First, the $\epsilon$-LDP guarantee on $\tilde{\mathbf{Y}}$ formally bounds how much information about individual genotypes is contained in the noised scores (Proposition~\ref{prop:sensitivity}).
Second, a server attempting to invert DP-noised embeddings to recover individual genotypes faces a severely underdetermined system: with $d \leq 10$ released components and $p \geq 28{,}000$ SNPs, the inversion problem has no unique solution, and DP noise further bounds reconstruction error in expectation. It is acknowledged that these bounds do not constitute cryptographic guarantees; the integration of secure multi-party computation is identified as a direction for future work.

\textbf{Scope of LDP guarantee.}
The $\epsilon$-LDP guarantee applies specifically to the released score vectors $\tilde{\mathbf{Y}} = \mathbf{Y} + \text{Lap}(\lambda)$. The PCA model $(\mathbf{V}, \boldsymbol{\mu})$ is shared under a separate trust assumption
and is not DP-protected in the current design.

\section{Proposed Framework}\label{sec:framework}
Before delving into the proposed framework, please refer to Table ~\ref{tab:symbols_notations} for a list of the symbols and notations used throughout the article.

Figure~\ref{fig:system_model}, summarizes the proposed pre-release auditing workflow. The researcher computes DP-noised PCA scores locally, the server maps public auxiliary datasets into the same PCA space, and membership-inference risk is estimated before publication.




    \textbf{Step~1 -- Data Collection, Local PCA, and DP Release:} The researcher collects genomic data and performs PCA locally to obtain a low-dimensional representation. The researcher computes PCA scores $\mathbf{Y} = (\mathbf{X} - \boldsymbol{\mu})\mathbf{V}_{1:d}$, applies the Laplace mechanism with noise scale $\lambda = 2\sqrt{d}/\epsilon$ (Proposition~\ref{prop:sensitivity}) to produce $\tilde{\mathbf{Y}} = \mathbf{Y} + \text{Lap}(\lambda)$, and shares only $\tilde{\mathbf{Y}}$ and the PCA model $(\mathbf{V}, \boldsymbol{\mu}, d)$ with the server.
    
    \textbf{Step~2 -- Data Sharing and Public Dataset Integration:} The researcher submits the DP-noised metadata and PCA model to the server. Simultaneously, the server gathers one or more relevant public $D_\text{ref}$ and projects them into the same coordinate system using the shared PCA model, ensuring that the transformation is consistent with the researcher's data. These public datasets serve as auxiliary sources to enhance risk estimation.
    
    \textbf{Step~3 -- Risk Quantification at the Server:} Once the metadata is received, the server undertakes a detailed analysis to quantify the associated privacy risk. Two key scenarios are considered:    
        \begin{enumerate}
            \item \textit{When sharing datasets:}
            \begin{enumerate}
                \item \textit{PCA-Based Analysis:} The server uses the researcher's shared PCA model to project one or more public datasets into the same low-dimensional coordinate system.
                \item \textit{Distance Computation:} Euclidean distances are calculated between the centroids of the researcher's data and those of the public datasets in the shared PCA space.
                \item \textit{Threshold Establishment and MIA Power Calculation:} A threshold $T$ is determined at the 5th percentile of distances from a designated control group within the public datasets, corresponding to a 5\% false-positive rate. MIA power is then computed as the fraction of the case group whose minimum distances to the researcher's data fall below $T$.
            \end{enumerate}

            \item \textit{When sharing statistics:} When only aggregated statistics (such as allele frequencies) are to be published, the server computes summary-level distance metrics by comparing the published statistics against corresponding metrics derived from public datasets. The Likelihood Ratio Test (LRT) is employed to quantify these differences and estimate the MIA risk from summary-level releases (see Section~\ref{sec:baseline-lrt}).

        \end{enumerate}
    \textbf{Step~4 -- Risk Reporting and Parameter Tuning:} 
    In both scenarios, the server reports the MIA power estimate and U-ReRo value to the researcher. This guides the researcher in adjusting the number of principal components, increasing the DP noise level (decreasing $\epsilon$), or refining the aggregation method. 


\section{Evaluation}\label{sec:evaluation}
For evaluation, the openSNP dataset~\cite{Opensnp} with 942 samples and 28,146 SNPs and the 1000~Genomes Project dataset~\cite{1000genom} with 600 samples and 37,751 SNPs  are used. Both datasets were preprocessed by removing SNPs with minor allele frequencies smaller than 0.01. The K-means clustering algorithm was applied to partition each dataset into distinct clusters ($k=5$ for openSNP,
$k=3$ for 1000~Genomes). These clusters were then used to simulate different case, control, and target datasets. The privacy budget $\epsilon$ is from 0.1 to 10 with a step size of 0.5. PCA dimensionality is varied using $d \in \{2, 3, 5, 7, 10\}$. All results are reported with 95\% confidence intervals derived from 10 independent trials with different random noise draws.

\subsection{Evaluation Metrics}
The Euclidean distance is used as the primary distance metric in this work. After applying PCA, the dataset is transformed into a lower-dimensional continuous numerical space, for which Euclidean distance provides an appropriate measure of the straight-line separation between points. Three complementary metrics are used to evaluate MIA risk. \textit{MIA Power}
measures the fraction of true dataset members correctly identified at a fixed 5\% false-positive rate. \textit{Precision and Recall} characterize the accuracy and completeness of the attack, respectively. \textit{U-ReRo}~\cite{Guerra2024Analysis} measures whether DP protections are applied symmetrically to members and non-members; values near zero indicate membership-neutral behavior. 

\subsection{Experimental Design}
Two scenarios are evaluated to assess the influence of auxiliary information quality. Each scenario represents a different level of auxiliary information available to the research follower, allowing assessment of privacy risk under varying conditions.

\begin{itemize}
    \item \textbf{Scenario~1, Target-Like Case-Control Group} \\
    The research follower has access to auxiliary data that closely resembles the researcher's dataset. This scenario evaluates how high auxiliary proximity influences MIA
  success and privacy risk.

    \item \textbf{Scenario~2, Non-Overlapping Case-Control Group} \\
    The research follower's auxiliary dataset does not overlap with the researcher's dataset. This scenario assesses the framework's behavior when the research follower's data is substantially different from the target dataset.
\end{itemize}



\subsection{Cluster Distances and Privacy Risk:}
Spatial proximity is quantified using Euclidean distances between the centroids of the target, case, and control datasets in the PCA-transformed space after z-score normalization of the PCA components. Intuitively, when auxiliary data is very close to or overlaps with the target distribution in this normalized
space, members and non-members become less separable under a distance-threshold decision rule, which lowers measured attack power at a fixed false-positive rate. Conversely, when auxiliary and target distributions are well separated, the same DP noise level becomes less effective at obscuring membership, leading to higher attack power.

\begin{figure*}[!t]
\centering
\setlength{\tabcolsep}{4pt}

\begin{tabular}{@{}cc@{}}
\includegraphics[width=.48\textwidth]{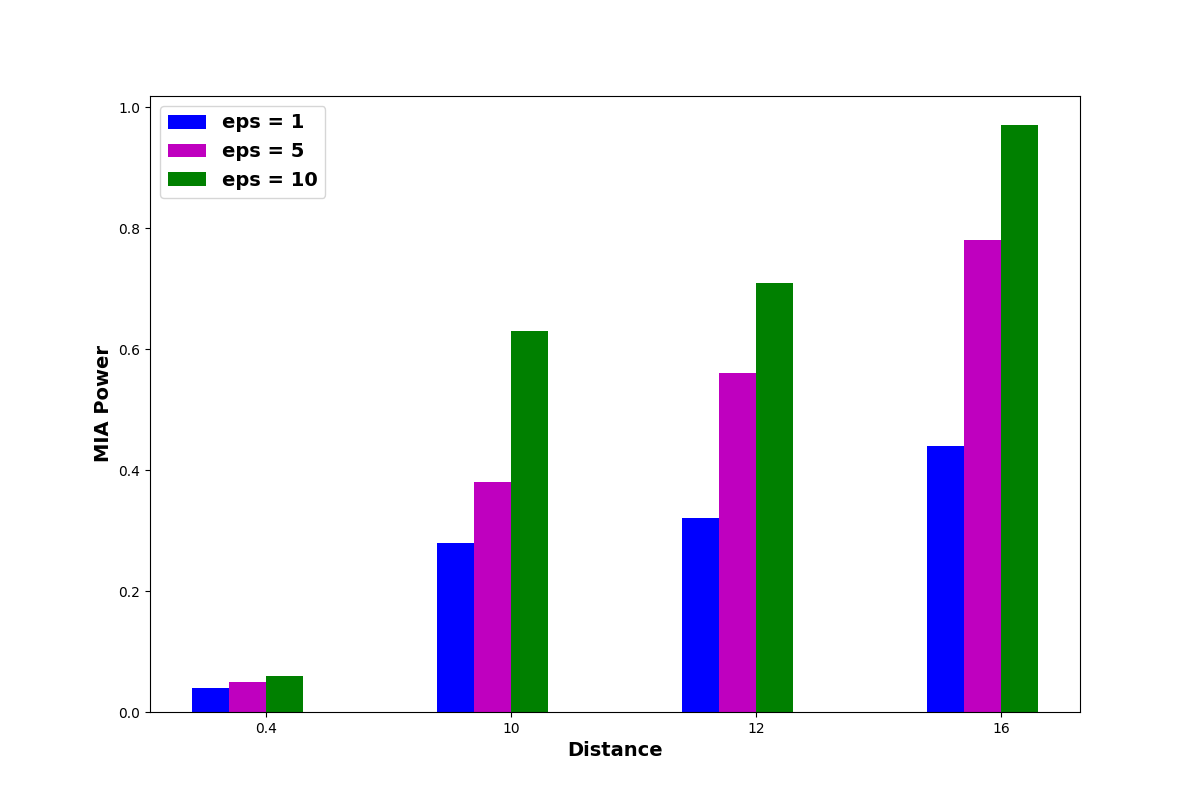} &
\includegraphics[width=.48\textwidth]{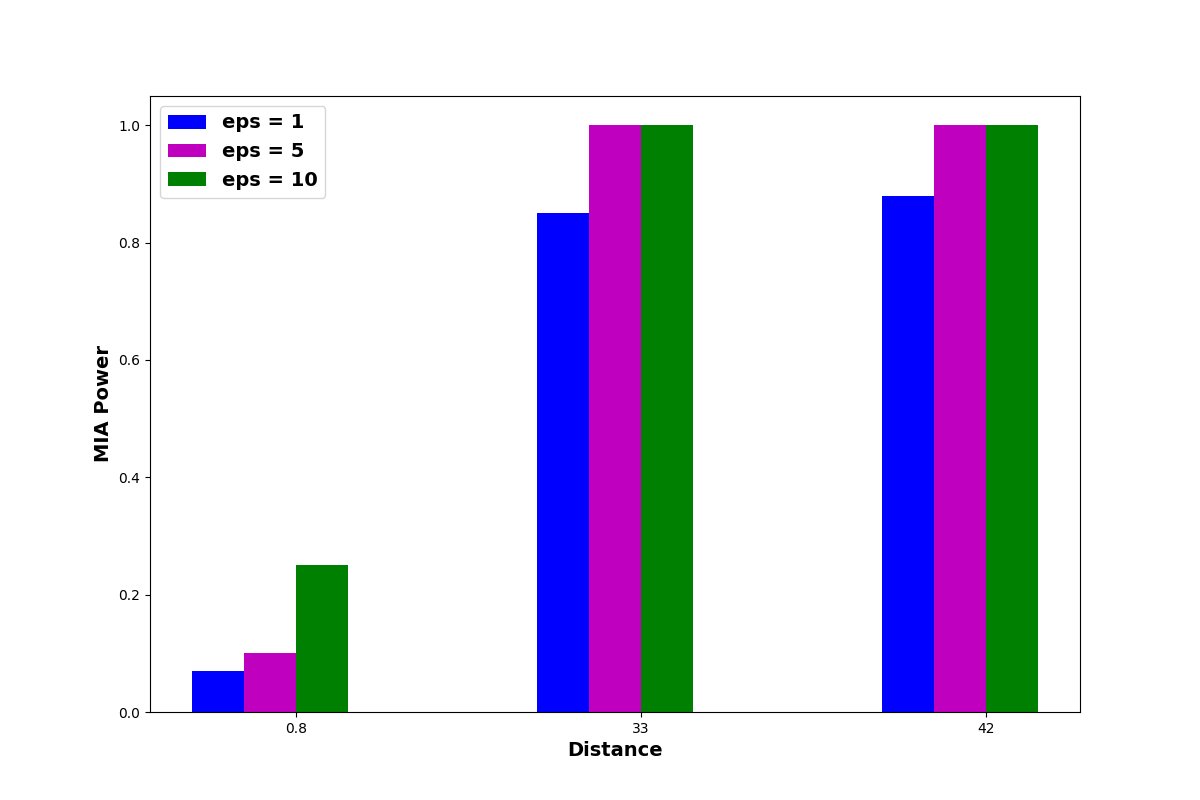} \\
\small (a) openSNP dataset &
\small (b) 1000 Genomes dataset
\end{tabular}

\caption{Membership inference attack (MIA) power under different distance thresholds and privacy budgets for the openSNP and 1000 Genomes datasets.}
\label{fig:power_distance}
\end{figure*}

Figure ~\ref{fig:power_distance} presents MIA power as a function of the distance between the researcher's PCA-transformed data and the public dataset, for $d=2$. Within each distance group, results are compared across different privacy budgets $\epsilon$. Note that $\epsilon$ defines how much noise the researcher adds to the PCA-transformed data before sharing it with the server: smaller $\epsilon$ means more noise (stronger privacy), while larger $\epsilon$ corresponds to less noise (weaker privacy). Here are key observations: 

MIA power increases with both centroid distance and $\epsilon$. Small normalized distances indicate substantial overlap between the researcher and auxiliary datasets, making members and non-members harder to distinguish. Larger distances create clearer separation and higher attack power. Similarly, larger $\epsilon$ adds less noise, increasing membership power and privacy risk.

\subsection{Privacy Metrics and Trends:}
\begin{figure*}[!t]
\centering
\setcounter{panel}{0}
\setlength{\tabcolsep}{3pt}

\begin{tabular}{@{}ccc@{}}

\refstepcounter{panel}\label{fig:openSNP_power}%
\includegraphics[width=.315\linewidth]{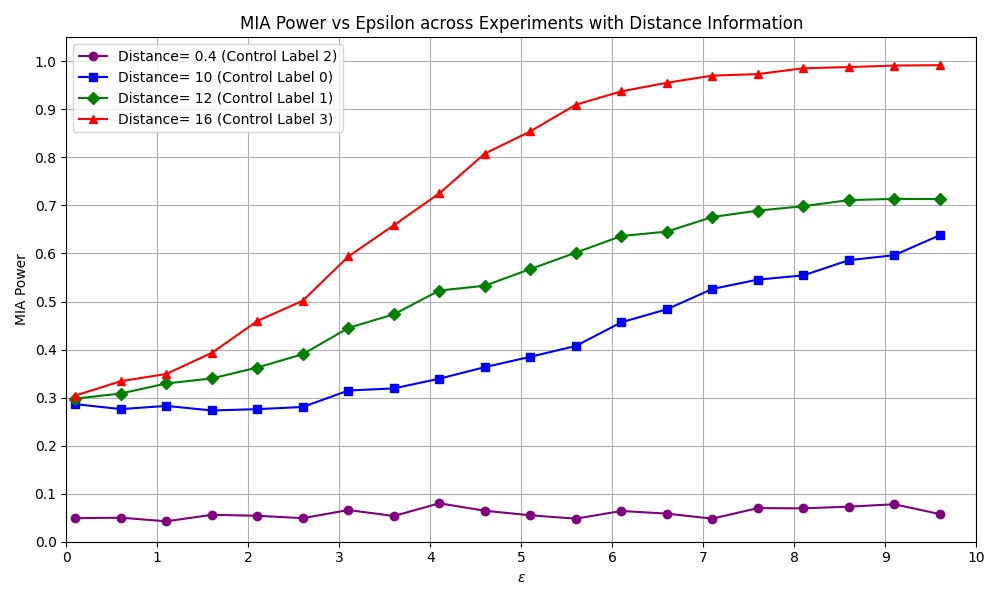} &
\refstepcounter{panel}\label{fig:openSNP_recall}%
\includegraphics[width=.315\linewidth]{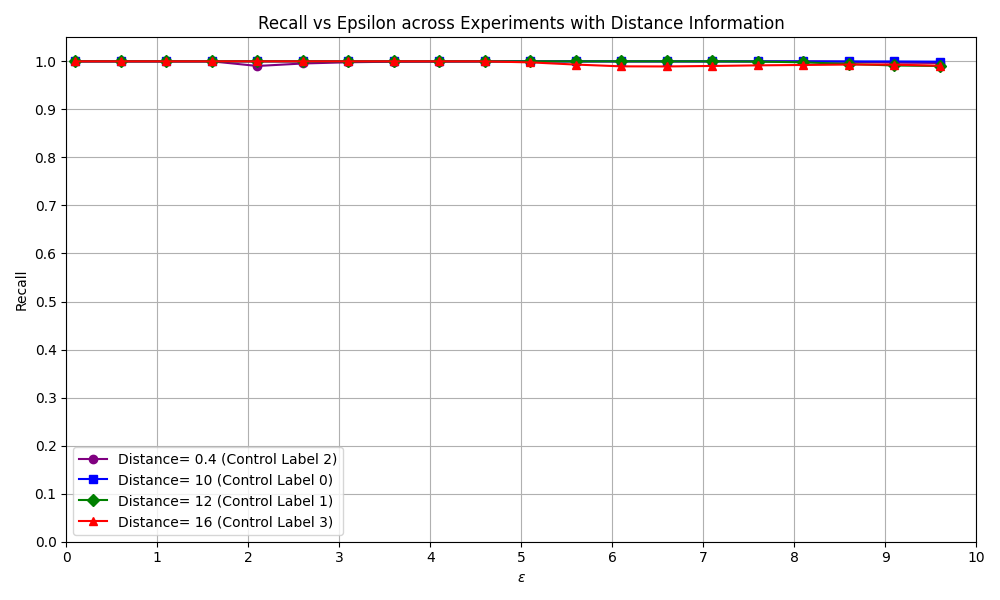} &
\refstepcounter{panel}\label{fig:openSNP_precision}%
\includegraphics[width=.315\linewidth]{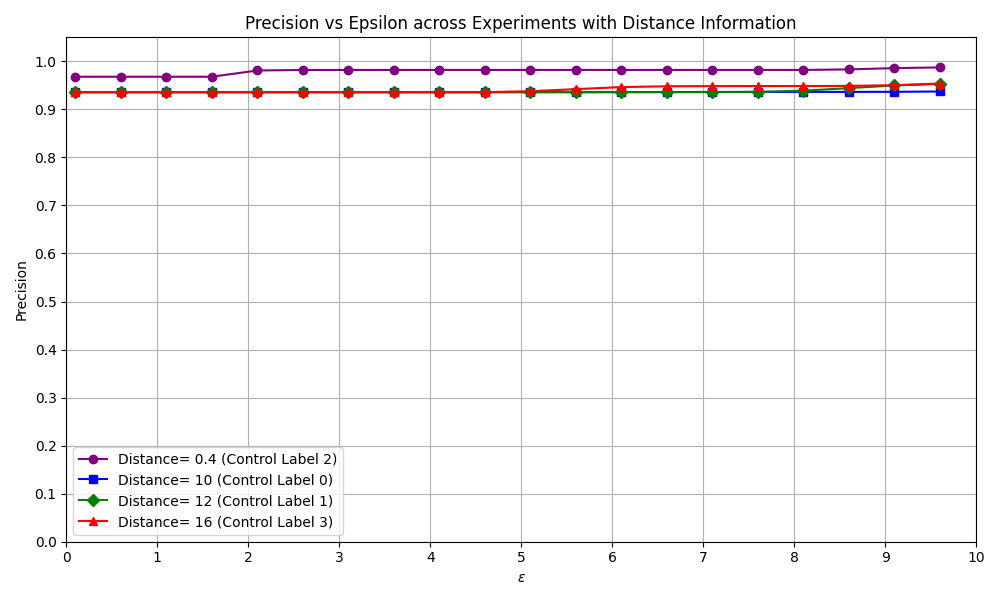} \\

\small (\ref{fig:openSNP_power}) Power vs.\ $\epsilon$ &
\small (\ref{fig:openSNP_recall}) Recall vs.\ $\epsilon$ &
\small (\ref{fig:openSNP_precision}) Precision vs.\ $\epsilon$ \\[0.8ex]

\refstepcounter{panel}\label{fig:1000_power}%
\includegraphics[width=.315\linewidth]{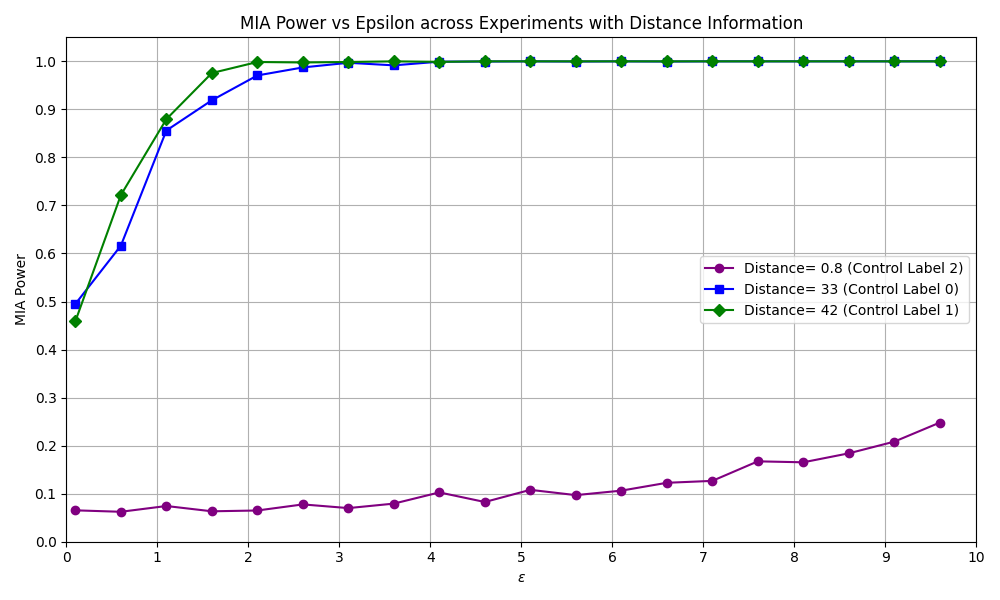} &
\refstepcounter{panel}\label{fig:1000_recall}%
\includegraphics[width=.315\linewidth]{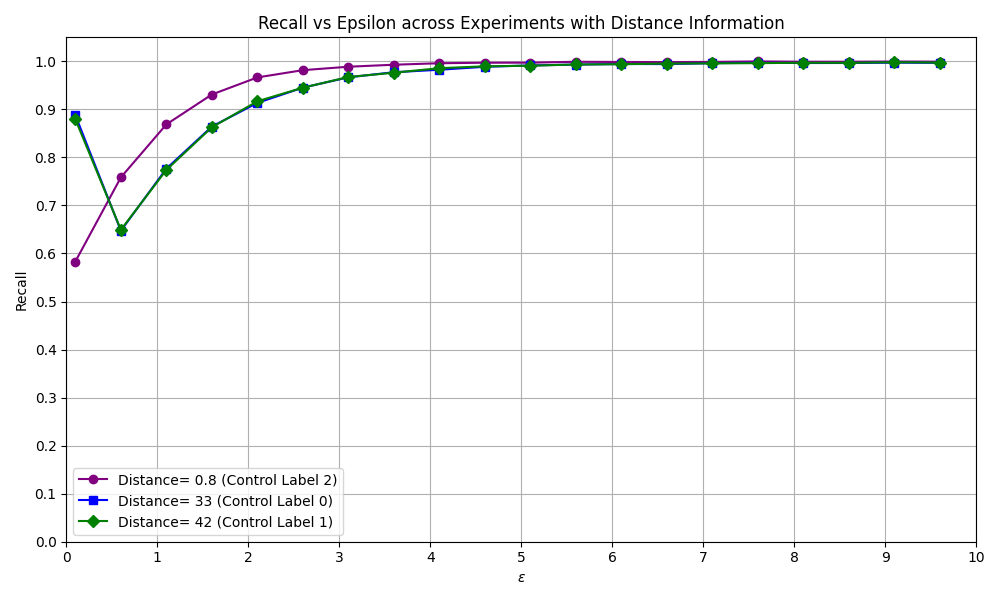} &
\refstepcounter{panel}\label{fig:1000_precision}%
\includegraphics[width=.315\linewidth]{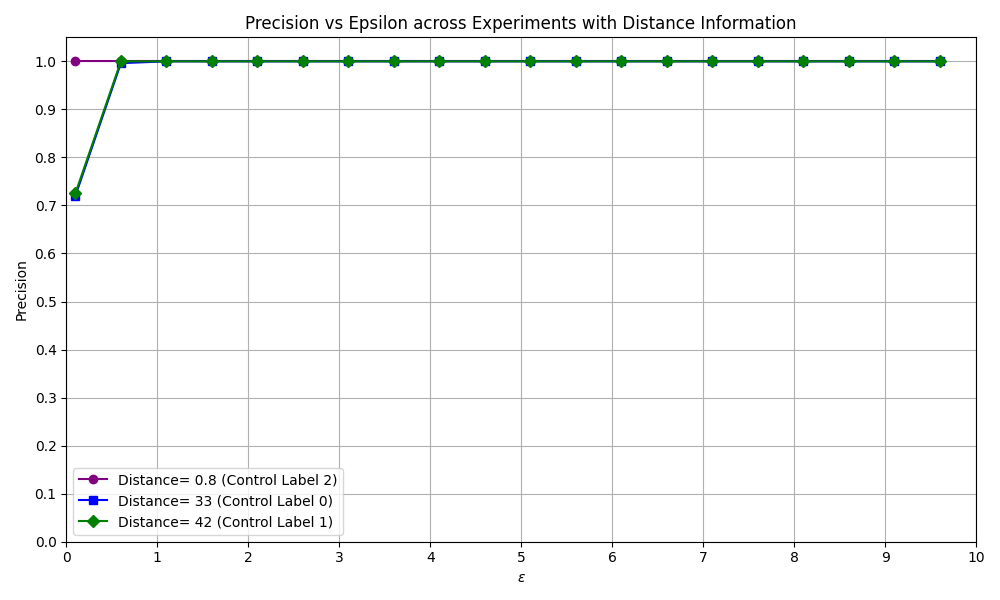} \\

\small (\ref{fig:1000_power}) Power vs.\ $\epsilon$ &
\small (\ref{fig:1000_recall}) Recall vs.\ $\epsilon$ &
\small (\ref{fig:1000_precision}) Precision vs.\ $\epsilon$

\end{tabular}

\caption{Impact of centroid distances on membership inference attack metrics under varying privacy budgets ($\epsilon$) for the openSNP and 1000 Genomes datasets. Panels (a--c) show power, recall, and precision for openSNP, respectively; panels (d--f) show the corresponding metrics for 1000 Genomes.}
\label{fig:combined_results}
\end{figure*}

Three metrics—power, precision, and recall—were evaluated under varying privacy budgets ($\epsilon$) and centroid distances:

\begin{enumerate}
    \item Power: As shown in Figures ~\ref{fig:combined_results}(\ref{fig:openSNP_power}) and ~\ref{fig:combined_results}(\ref{fig:1000_power}), power increases with higher $\epsilon$ values, indicating reduced privacy protection as noise decreases. For short centroid distances (e.g., less than~1 in normalized PCA space), power remains consistently low due to the ambiguity introduced by overlap. For larger distances, power is significantly higher, reflecting the vulnerability of separated datasets.
    
    \item Precision: Figures ~\ref{fig:combined_results}(\ref{fig:openSNP_precision}) and ~\ref{fig:combined_results}(\ref{fig:1000_precision}) illustrate that precision, or the research follower’s accuracy in correctly identifying members, increases with $\epsilon$ and is more pronounced for larger centroid distances. Notably, even for shorter distances,precision remains relatively stable, suggesting that ambiguity affects power more than precision.
    
    \item Recall: Recall trends, shown in Figures ~\ref{fig:combined_results}(\ref{fig:openSNP_recall}) and ~\ref{fig:combined_results}(\ref{fig:1000_recall}). Recall is more sensitive to dataset overlap than precision. Short distances result in fluctuating recall values, reflecting the challenge of distinguishing memberships when datasets are similar.
\end{enumerate}


\subsection{Baseline: Likelihood Ratio Test (LRT) on Released MAFs}
\label{sec:baseline-lrt}

We consider the standard membership-inference setting for GWAS summary-statistic release. A researcher releases minor allele frequencies (MAFs) for the top $L$ SNPs computed on a study cohort $D_R$ of size $n$. Given a target individual’s genotype and auxiliary ``population'' allele frequencies estimated from a reference cohort $D_P$, the adversary tests
\[
H_0:\; x \notin D_R \qquad \text{vs.} \qquad H_1:\; x \in D_R.
\]

\textbf{Dataset split and SNP selection.}
We instantiate this baseline on openSNP. The study cohort is constructed with $n=100$ individuals, balanced as 50 cases and 50 controls using a real openSNP dataset; the remaining openSNP individuals form the reference cohort $D_P$. To model common GWAS reporting practice, SNPs are ranked by GWAS significance computed within the study via an allelic $\chi^2$ test on the case/control labels, and the released statistics are the MAFs of the top-$L$ SNPs (smallest $p$-values), rather than random variants.

\textbf{Baseline attack and evaluation metric.}
Membership inference is performed using the LRT defined in Section~\ref{sec:background}. We report power (true-positive rate) at $\mathrm{FPR}=5\%$: the decision threshold $T$ is set so that $5\%$ of non-members (individuals from $D_P$) exceed $T$, and power is the fraction of members (individuals in $D_R$) whose scores exceed $T$.

\begin{table}[t]
\centering
\caption{LRT baseline on openSNP (study size $n=100$, 50 cases / 50 controls using the real data; reference = remaining individuals), power at $\mathrm{FPR}=5\%$. SNPs are selected by GWAS $p$-value computed on the study and released as MAFs.}
\label{tab:lrt-100}
\begin{tabular}{|ll|}
\hline
\textbf{\# released SNPs $L$} & \textbf{Power (Bernoulli LRT)} \\
\hline
10       & 0.15 \\
100      & 0.13 \\
1{,}000  & 0.39 \\
5{,}000  & 0.59 \\
10{,}000 & 0.60 \\
28{,}000 & 0.59 \\
\hline
\end{tabular}
\end{table}

This LRT baseline provides a standard reference point for membership risk under GWAS-style summary-statistic sharing. We use it to contextualize the privacy risk of our proposed approach, which instead shares DP-noised PCA outputs and explicitly models auxiliary-information quality (e.g., overlapping vs.\ non-overlapping scenarios) when estimating membership-inference risk.

Table~\ref{tab:lrt-100} shows that GWAS-style MAF releases can produce substantial membership-inference power at $\mathrm{FPR}=5\%$, reaching approximately $\approx 0.39$ at $L=1{,}000$ SNPs and $\approx 0.59$ at $L=5{,}000$ SNPs. Under target-like auxiliary conditions, the proposed DP-noised PCA framework yields lower MIA power across tested privacy budgets. However, when auxiliary data are well separated from the target, attack power can approach or exceed the LRT baseline at high $\epsilon$, emphasizing the need for dataset-specific governance thresholds.


\subsection{Unbiased Reconstruction Robustness (U-ReRo) and Membership Neutrality:}
As discussed in Section ~\ref{sec:background}, U-ReRo was implemented to evaluate the robustness of the proposed framework in ensuring unbiased protection against MIAs. 

In the evaluation, the parameters $\eta$ and $\gamma$ were set to~$0.5$, and the U-ReRo values consistently remained $\ll 0.1$ across all experimental conditions, confirming that the framework adheres to membership-neutral privacy guarantees. This result validates that adversarial success reflects only the information leaked due to dataset participation, ensuring unbiased and equitable privacy protection under the proposed DP mechanism.

\section{Discussion}
\label{sec:discuss}

\subsection{Summary of Findings}

The primary finding of this study is that geometric separation in PCA space between the researcher's dataset and public reference datasets is a dominant determinant of MIA vulnerability under the studied distance-threshold attack, rather than the perturbation level alone. When the researcher's dataset substantially overlaps with public auxiliary data, members and non-members become less separable in the shared PCA space and measured MIA power remains low. In contrast, when the researcher's dataset is well separated from available auxiliary data, membership inference becomes easier and higher attack power is observed. These findings underscore the importance of dataset-specific, pre-release risk assessment rather than universal perturbation recommendations.

\subsection{Relation to Current Practice}

The LRT baseline results demonstrate that sharing DP-noised PCA embeddings under overlapping auxiliary conditions yields risk estimates comparable to or
lower than the GWAS summary statistic releases currently permitted under NIH open-access policy (LRT power $\approx 0.39$--$0.59$ at $\ell = 1{,}000$--$5{,}000$ SNPs). Institutions seeking to implement quantitative pre-release privacy review may use these risk estimates to complement existing data-use agreement processes and IRB review protocols, particularly by evaluating dataset-specific membership inference risk before public or controlled-access release.



\subsection{Future Directions}

Three extensions are identified as priorities for future work. First, the
integration of cryptographic safeguards--specifically secure multi-party
computation protocols for the distance computation step--would eliminate the
server trust assumption and enable deployment in more adversarial institutional
environments. Second, biobank-scale validation on datasets of $n > 100{,}000$
individuals would establish whether the proximity-governed risk findings hold
at realistic GWAS scales and refine the governance thresholds accordingly.
Third, a multi-researcher extension enabling the server to aggregate risk
across concurrent studies would support cross-study privacy accounting, which
is increasingly necessary as public genomic repositories grow in size.
\section{Conclusion}\label{sec:conclusion}
This study offers a systematic evaluation of membership-inference risks in genomic data sharing, demonstrating how a proactive, pre-release framework can provide researchers and data governance bodies with actionable, dataset-specific guidance for privacy parameter selection. The proposed PCA/DP framework analyzes spatial relationships in genomic embeddings to explain why attacks succeed or fail; paired with MIA power, precision, recall, and U-ReRo, it clarifies how perturbation level and geometric separation jointly shape risk under a distance-threshold membership inference attack. The governance decision framework introduced here provides a quantitative complement to existing IRB review processes and NIH data-sharing guidelines. Looking ahead, the integration of cryptographic server safeguards, biobank-scale validation, and multi-researcher cross-study risk aggregation will further advance the goal
of principled, pre-release privacy auditing as a standard component of responsible genomic data sharing.


\bibliographystyle{IEEEtran}
{\bibliography{citation}}

@inproceedings{Wang2009Learning,
author = {Wang, Rui and Li, Yong Fuga and Wang, XiaoFeng and Tang, Haixu and Zhou, Xiaoyong},
title = {Learning your identity and disease from research papers: information leaks in genome wide association study},
year = {2009},
publisher = {Association for Computing Machinery},
address = {New York, NY, USA},
pages = {534–544},
numpages = {11},
location = {Chicago, Illinois, USA},
series = {CCS '09}
}

@article{Homer2008Resolving,
    author = {Homer, N and Szelinger, S and Redman, M and Duggan, D and Tembe, W and Muehling, J and Pearson, JV and Stephan, DA and Nelson, SF and Craig, DW} ,
    title = {Resolving individuals contributing trace amounts of DNA to highly complex mixtures using high-density SNP genotyping microarrays} ,
    journal = {PLoS Genetics},
    year = {2008} 
}

@inproceedings{chen2021Differential,
  title={Differential Privacy Protection Against Membership Inference Attack on Machine Learning for Genomic Data},
  author={Chen, Junjie and Wang, Wendy Hui and Shi, Xinghua},
  booktitle={Pacific Symposium on Biocomputing (PSB)},
  year={2021}
}

@article{zhang2024bayesnash,
  title={Bayes-Nash Generative Privacy Protection Against Membership Inference Attacks},
  author={Zhang, Tao and Venkatesaraman, Rajagopal and De, Rajat K. and Malin, Bradley A. and Vorobeychik, Yevgeniy},
  journal={arXiv preprint arXiv:2410.07414},
  year={2024}
}

@inproceedings{humbert2017Quantifying,
author={M. Humbert and E. Ayday and J.-P. Hubaux and A. Telenti},
journal={ACM Transactions on Privacy and Security}, 
title={Quantifying Interdependent Risks in Genomic Privacy},
year={2017},
volume={20},
number={1},
pages={1-31},
ISSN={2471-2566},
month={February}
}

@inproceedings{Yilmaz2022Genomic,
author = {Yilmaz, Emre and Ji, Tianxi and Ayday, Erman and Li, Pan},
title = {Genomic Data Sharing under Dependent Local Differential Privacy},
year = {2022},
publisher = {Association for Computing Machinery},
address = {New York, NY, USA},
doi = {10.1145/3508398.3511519},
booktitle = {Proceedings of the Twelfth ACM Conference on Data and Application Security and Privacy},
pages = {77–88},
numpages = {12},
location = {Baltimore, MD, USA},
series = {CODASPY '22}
}

@article{Bu2021Haplotype,
    author = {Bu, Diyue and Wang, Xiaofeng and Tang, Haixu},
    title = {Haplotype-Based Membership Inference from Summary Genomic Data},
    journal = {Bioinformatics},
    volume = {37},
    number = {Supplement 1},
    pages = {i161--i168},
    year = {2021},
    month = {07},
    issn = {1367-4803},
    doi = {10.1093/bioinformatics/btab305}
}

@inproceedings{costello2015geppetto,
  author    = {Costello, Craig and et al.},
  title     = {Geppetto: Versatile verifiable computation},
  booktitle = {Proc. IEEE Symp. Secur. Privacy (SP)},
  year      = {2015}
}

@inproceedings{davidson2011provenance,
  author    = {Davidson, Susan and Khanna, Sanjeev and Milo, Tova and Panigrahi, Debmalya and Roy, Sudip},
  title     = {Provenance views for module privacy},
  booktitle = {Proc. 30th ACM SIGMOD-SIGACT-SIGART Symp. Principles Database Syst. (PODS)},
  year      = {2011}
}

@inproceedings{davidson2011on,
  author    = {Davidson, Susan and Khanna, Sanjeev and Roy, Sudip and Stoyanovich, Julia and Tannen, Val and Chen, Yi},
  title     = {On provenance and privacy},
  booktitle = {Proc. 14th Int. Conf. Database Theory (ICDT)},
  year      = {2011}
}

@inproceedings{Guerra2024Analysis,
    author = {Guerra-Balboa, Patricia and Sauer, Annika and Strufe, Thorsten},
    title = {Analysis and Measurement of Attack Resilience of Differential Privacy},
    year = {2024},
    isbn = {9798400712395},
    publisher = {Association for Computing Machinery},
    doi = {10.1145/3689943.3695046},
    booktitle = {Proceedings of the 23rd Workshop on Privacy in the Electronic Society},
    pages = {155--171},
    numpages = {17}
}

@misc{Opensnp,
  author       = {{openSNP}},
  title        = {{openSNP}},
  howpublished = {\url{https://opensnp.org/}},
  note         = {Accessed: 2025-04-20}
}

@misc{dbGap,
    author       = {{dbGap}},
    title = {{The Database of Genotypes and Phenotypes (dbGaP)}},
    howpublished = {\url{https://www.ncbi.nlm.nih.gov/gap/}}
}

@misc{ UK-BioBank,
author       = {{The UK-BioBank}},
  title     = {{UK BioBank}},
  howpublished = {\url{https://www.ukbiobank.ac.uk/}}
}

@misc{1000genom,
  author       = {{The 1000 Genomes Project}},
  title        = {{1000 Genomes Project}},
  howpublished = {\url{https://www.internationalgenome.org/}},
  note         = {Accessed: 2025-01-20}
}

@article{Halimi2022Privacy,
    author = {Halimi, A and Dervishi, L and Ayday, E and Pyrgelis, A and Troncoso-Pastoriza, JR and Hubaux, JP and Jiang, X and Vaidya, J},
    title = {Privacy-Preserving and Efficient Verification of the Outcome in Genome-Wide Association Studies},
    journal = {Proc Priv Enhanc Technol},
    year = {2022}
}

@article{Abdi2010Principal,
author = {Abdi, Hervé and Williams, Lynne J.},
title = {Principal component analysis},
journal = {WIREs Computational Statistics},
volume = {2},
number = {4},
pages = {433-459},
doi = {https://doi.org/10.1002/wics.101},
year = {2010}
}

@article{Pedregosa2011Scikit,
  author  = {Fabian Pedregosa and Ga{{\"e}}l Varoquaux and Alexandre Gramfort and Vincent Michel and Bertrand Thirion and Olivier Grisel and Mathieu Blondel and Peter Prettenhofer and Ron Weiss and Vincent Dubourg and Jake Vanderplas and Alexandre Passos and David Cournapeau and Matthieu Brucher and Matthieu Perrot and {{\'E}}douard Duchesnay},
  title   = {Scikit-learn: Machine Learning in Python},
  journal = {Journal of Machine Learning Research},
  year    = {2011},
  volume  = {12},
  number  = {85},
  pages   = {2825--2830}
}

@InProceedings{Dwork2008Differential,
author="Dwork, Cynthia",
title="Differential Privacy: A Survey of Results",
booktitle="Theory and Applications of Models of Computation",
year="2008",
publisher="Springer Berlin Heidelberg",
address="Berlin, Heidelberg",
pages="1--19",
isbn="978-3-540-79228-4"
}

@InProceedings{Dwork2006Calibrating,
author="Dwork, Cynthia
and McSherry, Frank
and Nissim, Kobbi
and Smith, Adam",
editor="Halevi, Shai
and Rabin, Tal",
title="Calibrating Noise to Sensitivity in Private Data Analysis",
booktitle="Theory of Cryptography",
year="2006",
publisher="Springer Berlin Heidelberg",
address="Berlin, Heidelberg",
pages="265--284"
}

@INPROCEEDINGS{Duchi2013Local,
  author={Duchi, John C. and Jordan, Michael I. and Wainwright, Martin J.},
  booktitle={2013 IEEE 54th Annual Symposium on Foundations of Computer Science}, 
  title={Local Privacy and Statistical Minimax Rates}, 
  year={2013},
  volume={},
  number={},
  pages={429-438},
  doi={10.1109/FOCS.2013.53}
}

@inproceedings{Kairouz2014Extremal,
 author = {Kairouz, Peter and Oh, Sewoong and Viswanath, Pramod},
 booktitle = {Advances in Neural Information Processing Systems},
 publisher = {Curran Associates, Inc.},
 title = {Extremal Mechanisms for Local Differential Privacy},
 year = {2014}
}

@article{Zerhouni2008Protecting,
author = {Elias A. Zerhouni  and Elizabeth G. Nabel },
title = {Protecting Aggregate Genomic Data},
journal = {Science},
volume = {322},
number = {5898},
pages = {44-44},
year = {2008}}

@article{Couzin2008Whole-Genome,
author = {Jennifer Couzin },
title = {Whole-Genome Data Not Anonymous, Challenging Assumptions},
journal = {Science},
volume = {321},
number = {5894},
pages = {1278-1278},
year = {2008}
}

\end{document}